\documentstyle[aps,epsfig,floats]{revtex}
\newcommand {\be} {\begin{equation}}
\newcommand {\ee} {\end{equation}}
\newcommand {\Be} {\begin{eqnarray*}}
\newcommand {\Ee} {\end{eqnarray*}}
\newcommand {\bey}{\begin{eqnarray}}
\newcommand {\eey}{\end{eqnarray}}
\newcommand {\oti}{\tilde\omega}

\begin{document}
\begin{center}

\Large{\bf On the anomalous thermal conductivity\\
of one-dimensional lattices}\\

\vspace{0.5cm}

{\large Stefano Lepri}
\footnote{also Istituto Nazionale di Fisica della Materia,
Unit\`a di Firenze} \\
{\small\it
Max-Planck-Institut f\"ur Physik Komplexer Systeme,
N\"othnitzer Str. 38, D-01187 Dresden, Germany\\
{\rm lepri@mpipks-dresden.mpg.de}\\
}

\vspace{0.3cm}
{\large Roberto Livi} $^{*\dagger}$\\
{\small\it
Dipartimento di Fisica dell'Universit\`a,
L.go E.Fermi 2 I-50125 Firenze, Italy\\
{\rm livi@fi.infn.it}\\
}

\vspace{0.3cm}
{\large Antonio Politi} \footnote{
also Istituto Nazionale di Fisica Nucleare, Sezione di Firenze }\\
{\small\it
Istituto Nazionale di Ottica,
L.go E.Fermi 6 I-50125 Firenze, Italy\\
{\rm politi@ino.it}\\
}

\end{center}

\date{\today}
\begin{abstract}
{The divergence of the thermal conductivity in the thermodynamic limit is
thoroughly investigated. The divergence law is consistently determined with
two different numerical approaches based on equilibrium and non-equilibrium
simulations. A possible explanation in the framework of 
linear-response theory is also presented, which traces back the physical origin of this 
anomaly to the slow diffusion of the energy of long-wavelength Fourier modes. Finally,  
the results of dynamical simulations are compared with the predictions of
mode-coupling theory.}

\vspace{0.4cm}
{\it To appear in Europhysics Letters (1998)}\\
\end{abstract}

\vspace{1.0cm}

An extremely idealized, but physically meaningful, model of an insulating
solid is a set of $N$ atoms of mass $m$, arranged on a 1-d lattice with 
spacing $a$ and coupled by nonlinear forces. Denoting with $q_l$ the
displacement of the $l$-th particle from its equilibrium position $la$, the
corresponding Hamiltonian reads
\be
H = \sum_{l=1}^N \left[{p_l\over 2m}^2+ V(q_{l+1}-q_{l})\right] \quad ,
\label{hami}
\ee
where $p_l=m\dot q_l$. If the chain is put in contact at its boundaries
with two heat baths at different temperatures $T_0$ and $T_0+\Delta T$, a
nonequilibrium stationary state arises, characterized by a non vanishing
average heat flux $J$. The microscopic definition of $J$ is
\cite{choq,noi2}
\be
J \;=\; {1\over N} \sum_l j_l = {a\over 2}\,{1\over N} \sum_l
\dot q_l\, \left(f_{l+1}+f_l\right)
\label{global}
\ee
where $f_l=-V'(q_l-q_{l-1})$ is the interaction force and $j_l$ represents
the local flux at site $l$ (i.e. the sum of the fluxes of potential energy 
from its neighbours). The lattice thermal conductivity $\kappa$ can be defined 
through the Fourier law,
\begin{equation}
\langle J \rangle  = -\kappa {dT\over dx} \quad ,
\label{conduc}
\end{equation}
where $x=la$ is the coordinate along the chain and $\langle \cdot
\rangle$ denotes a time average in the stationary regime.

The task of statistical mechanics is to explain such a macroscopic law,
starting from the microscopic interactions. Long time ago, Peierls
proposed a theory based on the phonon scattering mechanism in the perturbative 
and close-to-equilibrium limits \cite{peierls}. The common belief was thereby 
that a microscopic theory able to include the effects of disorder and/or 
nonlinearity should ensure the validity of Eq.~(\ref{conduc}) also in the 
case of strong anharmonicity and for systems far from equilibrium. 

Nevertheless, the rigorous treatment of isotopically disordered harmonic 
chains showed that disorder alone is not sufficient for obtaining 
a reliable model: the thermal conductivity diverges as $\kappa \propto N^{1/2}$ 
\cite{papa}. This result enforced the necessity to verify whether nonlinearities
alone can eliminate such an anomaly. Recently, the problem has been reconsidered
by resorting to nonequilibrium molecular-dynamics simulations. Among the many 
attempts, the most convincing results in favour of a finite $\kappa$ were obtained 
only for rather artificial systems, namely the ding-a-ling models \cite{casa,pros}.

For more realistic systems, like the diatomic Toda chain a finite conductivity
was found for small enough mass ratio \cite{jack}, while the analysis of model 
(\ref{hami}) with the Fermi-Pasta-Ulam (FPU) potential
\be
V(y)\;=\; {1\over 2}m\omega_0^2y^2 \,+\,{1\over 4}g y^4 \quad,
\label{fpu}
\ee
gives evidence of a power-law divergence $\kappa \propto N^{\beta}$ for 
$N\gg 1$ \cite{noi,noi2}. We have verified that this result is robust against 
changes in the number of thermostatted atoms, in the boundary conditions and in 
the type of heat baths (time-reversible as well as stochastic ones). More 
precisely, simulations reported in \cite{noi2} with a large temperature gradient 
($\Delta T = 128$ and $T_0 = 24$) reveal a crossover from $\beta \approx 0.45$ 
for $N < 250$ to $\beta \approx 0.38$ for larger $N$, while new simulations made 
with $\Delta T = 16$ and $T_0 = 80$ (Fig.~1) yield $\beta = 0.37$ up to $N=2048$. 

For sufficiently small $\Delta T$ and large $N$, the system is close to an equilibrium 
state at temperature $T = T_0 + {{\Delta T}\over{2}}$. One can then rely on 
the Green-Kubo formula \cite{kubo}
\be
\label{gk}
\kappa = {1\over k_BT^2} \, \int_0^\infty \, C_J(t) dt
\ee
where $C_J(t) = N \langle J(t)J(0)\rangle $ is the correlation function
at equilibrium. We have numerically computed $C_J(t)$ for the FPU chain with 
Born-Von Karman boundary conditions, and initial conditions sampled according 
to the canonical Gibbs measure. In Fig.~2, we report the power spectrum 
$S(\omega)$ of the heat flux $J$ for $N=512$. In the low-frequency region
there is a power-law divergence as $\omega^{-0.37}$, which
corresponds to a decay of $C_J(t)$ as $t^{-0.63}$. 

A crucial information for comparing these results with those obtained with 
heat baths comes from the spatio-temporal correlation function 
$C_j(l,t) = \langle j_l(t)j_0(0)\rangle$ of the local flux. Indeed, the peaks 
shown in Fig.~3 indicate that energy propagates at constant
velocity $v_p$. Therefore, one can estimate the dependence of $\kappa$ on $N$
from the asymptotic behaviour of $C_J(t)$, by restricting the integral in
(\ref{gk}) to times smaller than the ``transit time'' $N/v_{p}$. This amounts 
to ignoring all the contributions from sites at a distance larger than $N$. 
With the above estimate of $C_J$, one obtains that 
$\kappa \propto N^{0.37}$, in agreement with the direct estimate obtained from 
the non-equilibrium simulations in Fig.~1.

Despite the remarkable consistency of the numerical results, the scenario is 
quite unsatisfactory on a theoretical ground, especially because there is 
neither a clue about the very reason of the observed anomalous transport, nor 
an explicit estimate of the measured exponent. In the following, we present 
a possible explanation based on a kind of hydrodynamic approach with 
long-wavelength modes interacting with a stochastic-like reservoir. 

In fact, it has been observed \cite{alabi} that even at high energies, where
the FPU-system is definitely nonintegrable, the energy of its long-wavelength 
Fourier modes diffuses very slowly (with respect to their period). Accordingly,
they may effectively act as undamped transport channels, a feature that
reflects itself in the relaxation of fluctuations as measured by $C_J(t)$.
Our goal is now to obtain an estimate of the asymptotic behaviour of $C_J(t)$, 
by describing the relaxation close to equilibrium of the Fourier-mode 
amplitudes $A_k$ and $B_k$, defined by 
\be
q_l = {1\over\sqrt{N}}\sum_{k=1}^{N/2}\left[
A_k \cos\left({2\pi ka\over N}l\right)+ 
B_k \sin\left({2\pi ka\over N}l\right)\right] \quad ,
\label{modi}
\ee
where we have made the assumption that the zero-mode $A_0$ is identically 
zero. The customary approach would be to write stochastic equations for $A_k$ 
and $B_k$. The general strategy involves projection on the subspace of such 
(slow) variables and yields linear Langevin equations with memory terms 
\cite{kubo}. If we assume that a separation of proper time scales is possible, 
the former reduce to Markovian equations \cite{lepri},
\be
\label{markov}
 \ddot A_k + \gamma_k \dot A_k + {\oti_k}^2 A_k = \xi_k \quad;\quad
 \ddot B_k + \gamma_k \dot B_k + {\oti_k}^2 B_k = \eta_k 
\ee
where $\gamma_k$ are phenomenological relaxation rates, $\xi_k$ and $\eta_k$ 
are mutually independent, Gaussian, random and white processes \cite{note1}. 
Nonlinear effects are taken into account by renormalization of the bare 
dispersion relation as \cite{alabi}
\be
\tilde\omega_k \;=\; \alpha\;\; \omega_k =
\alpha \,\cdot 2 \omega_0 \sin\left({\pi ka\over N}\right) \quad ,
\label{omtilde}
\ee
i.e. by renormalizing the sound velocity from $v=a\omega_0$ to
$\tilde v = \alpha \, v$, where, for the parameters of Fig.~1, 
$\alpha = \sqrt{[1 + 2(1+0.72\,\varepsilon)^{1/2}]/3}$ and
$\varepsilon(T)$ is the internal energy \cite{alabi}.
 
A first confirmation that transport is basically determined by the low $k$-modes 
comes from the observation that $\tilde v$ is remarkably close to $v_p$: for the 
parameter values considered in Fig.~3, $\tilde v = 2.38$, while $v_{p}=2.47$.

Let us now split $V$ into its harmonic and anharmonic parts and consequently
write the flux (\ref{global}) as $J=J_H+J_A$. From Eq.~(\ref{modi}) one
obtains
\be
J_H \;=\; {m\over N} \sum_{k=1}^{N/2} c_k \omega_k 
\left(A_k \dot B_k - \dot A_k B_k\right) \quad ,
\label{jh}
\ee
where $c_k = d\omega_k/dk$ is the bare phase velocity. 
For a strongly anharmonic system, like the one we consider here, we do 
not expect $J_A$ to be negligible. Nevertheless, we shall now argue that 
the leading asymptotic behaviour of $C_J$ can be estimated by computing
the autocorrelation of $J_H$ alone. We limit ourselves to the FPU case (\ref{fpu}), 
for which $J_A$ is a sum of fourth order terms coupling modes with indices 
$k,k',k'',k'''$. By virtue of Eq.~(\ref{markov}), each low-$k$ component 
oscillates approximatively at a frequency that is an algebraic sum of 
$\oti_{k}$, $\oti_{k'}$, $\oti_{k''}$ and $\oti_{k'''}$. It is then reasonable 
to assume that rapidly oscillating terms will be relevant only for short times, 
while the asymptotic dominant contributions originate from the resonant terms.
This fact, together with selection rules, require that only 
terms fulfilling the constraints
\bey
\label{selrul}
& k+k'+k''+k''' \;=\; 0 \\
& \tilde\omega_{k}+\tilde\omega_{k'}-\tilde\omega_{k''}-\tilde\omega_{k'''}
\;=\;0
\nonumber
\eey
must be retained. For small $k$, we can linearize the dispersion
relation in the second of Eqs.~(\ref{selrul}) obtaining
\be
J_A \approx J_H\,{g\over m \omega_0^4}{1\over N} 
\sum_{k} \, \omega_k^2 \left(A_k^2 + B_k^2 \right) \quad .
\label{japprox}
\ee
In order to evaluate the contribution of $J_A$ to $C_J$, let us first notice 
that the former is the product of two time-dependent terms: $J_H$ and the sum
${\cal S} = \sum \omega_k^2(A_k^2+B_k^2)$. Any given $k$-contribution
in the expression of $J_H$ (see Eq.~(\ref{jh})), is correlated with only one 
(with the same $k$) out of the $N$ addenda of the sum in Eq.~(\ref{japprox}).
Accordingly, in the thermodynamic limit, we can consider $\cal S$ as 
uncorrelated with $J_H$.
Moreover, since $\cal S$ is the sum of uncorrelated positive terms,
we can substitute it with its average value and write
\be
J\;\approx\; J_H \left(1+{4g\over m^2 \omega_0^4}{\cal U}(T)\right)
\label{jfinale} \quad ,
\ee
where we have also taken into account energy equipartition in its 
generalized form, i.e. $m\omega_k^2 \langle A_k^2\rangle = 
m\omega_k^2 \langle B_k^2\rangle = {\cal U}(T)$, for some function 
${\cal U}$ independent of $k$. We numerically checked the reliability of 
our approximations by first comparing expression (\ref{japprox}) with $J_A$ 
and then by verifying that the autocorrelation functions of $J_H$ and $J$ 
are proportional. 

Let us compute the autocorrelation of $J_H$. Using Eqs.~(\ref{markov}), one 
finds that the new variables $W_k=A_k \dot B_k - \dot A_k B_k$ satisfy the 
Langevin equation
\be
\dot W_k \;=\; -\gamma_k \,W_k + \zeta_k \quad .
\ee
In the limit of small $\gamma_k$, $\zeta_k$ is a Gaussian and delta-correlated
random process, so that for large $N$ 
\be
\langle  J_H(t)J_H(0) \rangle = 
{4\over N^2} k_BT {\cal U}(T) \sum_k {c_k}^2 e^{-\gamma_k t} 
\approx  {1\over N}
{2a\over \pi}k_BT {\cal U}(T)\int_0^{\pi/a} dk \, c^2(k) \, e^{-\gamma(k) t}
\quad .
\label{jj}
\ee
Due to the conservation of total momentum, one expects $\gamma_k$ to vanish 
for vanishing $k$. By drawing an analogy with the known nonanalytic behaviour
of dispersion relations of low-dimensional fluids \cite{thanks}, we may
assume $\gamma(k) \approx \nu k^\mu$, to leading order in $k$ \cite{pomeau}. 
We also
suppose that $\mu>1$ consistently with the observed fact that dissipation 
occurs on time scales much longer than the proper period. Since the conduction 
properties are determined by low-$k$ modes, we can approximate Eq.(\ref{jj})
by expanding $c(k)$ around $k=0$ so that
\be
\langle  J_H(t)J_H(0) \rangle \propto {1\over N}
{ v^2 k_BT{\cal U}(T) a \over (\nu t)^{1\over \mu}} \,\left[ 1
+ {\cal O} \left(t^{-{2\over \mu}} \right)\right].
\label{corre}
\ee
The power-law decay of $C_J(t)$ implies that $\kappa$ should diverge as 
$N^{1-1/\mu}$. 

The key point is then the evaluation of $\mu$. In the case of fluids, this is 
generally accomplished by resorting to a self-consistent mode-coupling approximation 
\cite{pomeau}. In the present context, it is reasonable to conjecture,  on the basis 
of Eqs.~(\ref{markov}), that the behaviour of low-$k$ modes can be assimilated to 
that of hydrodynamic modes in a 1d fluid, at least at such high temperatures 
(see \cite{lepri} for a more detailed analytical discussion). One 
should then expect $\mu=5/3$ \cite{ernst}, thus yielding $C_J(t)\propto t^{-3/5}$ 
and $\kappa \propto N^{2/5}$. This prediction is compared with the numerical results 
in the inset of Fig.~2. The discrepancy with the best-fit value (0.37) can be 
definitely attributed to finite-size effects. As long as the above conjecture proves 
correct, our numerical results provide the most striking confirmation of the mode 
coupling predictions.

The generality of our arguments suggests that any nonlinear 1-d model 
with an acoustic spectral component should exhibit the same anomaly. It 
has however to be admitted that the mentioned Toda model in which an 
optical branch is also present \cite{jack} is qualitatively different. 

Finally, we observe that there is no reason for expecting the same phenomenon 
not to occur also in higher dimensions, independently in each acoustic
branch. On the other hand, the exponent $\mu$ may well depend on $d$ and,
moreover, the integral in Eq.~(\ref{jj}) has to be weighted with the phononic 
density of states, which, for small $k$, vanishes as $k^{d-1}$. One 
can conjecture that such a mechanism does not introduce any divergence in the 
conductivity for $d = 3$, while a logarithmic divergence should be 
observed for $d = 2$. This prediction provides a unified explanation for both 
the anomalous transport of 1-d systems and the finite conductivity of 3-d lattices.

\begin{figure}
\vspace{-2cm}
\hspace{.7cm}\centering\epsfig{figure=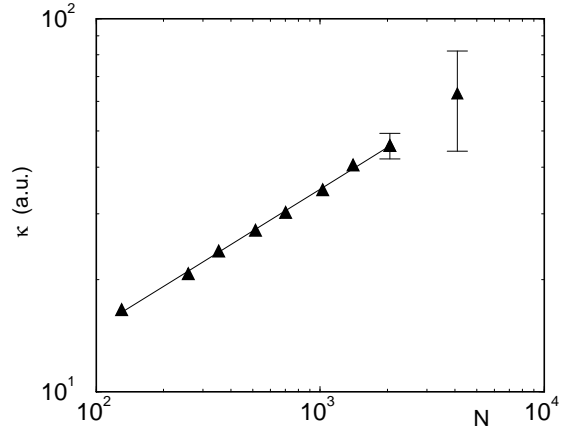,width=8cm}
\begin{minipage}[b]{15cm}
\caption{\baselineskip=12pt
Divergence of the thermal conductivity with the system size for the
FPU chain ($m=\omega_0=1$, $g=0.1$) as obtained from nonequilibrium
molecular dynamics (see Refs. \protect\cite{noi,noi2} for details).
Statistical errors are displayed only when larger than the symbols' size.
}
\end{minipage}
\end{figure}

\begin{figure}[h]
\vspace{-1cm}
\hspace{.7cm}\centering\epsfig{figure=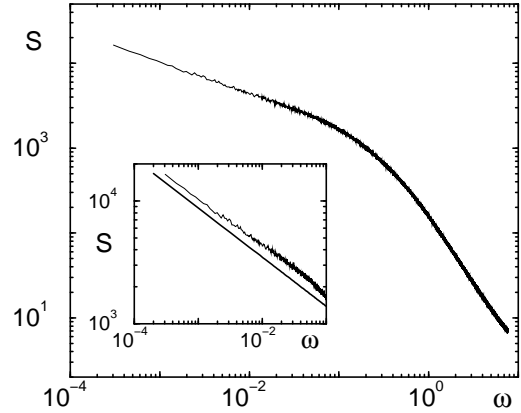,width=8cm}
\begin{minipage}{15cm}
\caption{\baselineskip=12pt
Power spectrum $S(\omega)$ (in arbitrary units) of the global flux for the
FPU chain (same parameters as in Fig.~1) with $N=1024$ at temperature $T=110.7$.
The curve results from an average over 1400 independent initial conditions.
A blow-up of the low-frequency region is reported in the inset:
the straight line corresponds to the $1/\omega^{2/5}$ behaviour.}
\end{minipage}
\end{figure}

\begin{figure}
\vspace{-2cm}
\hspace{.7cm}\centering\psfig{figure=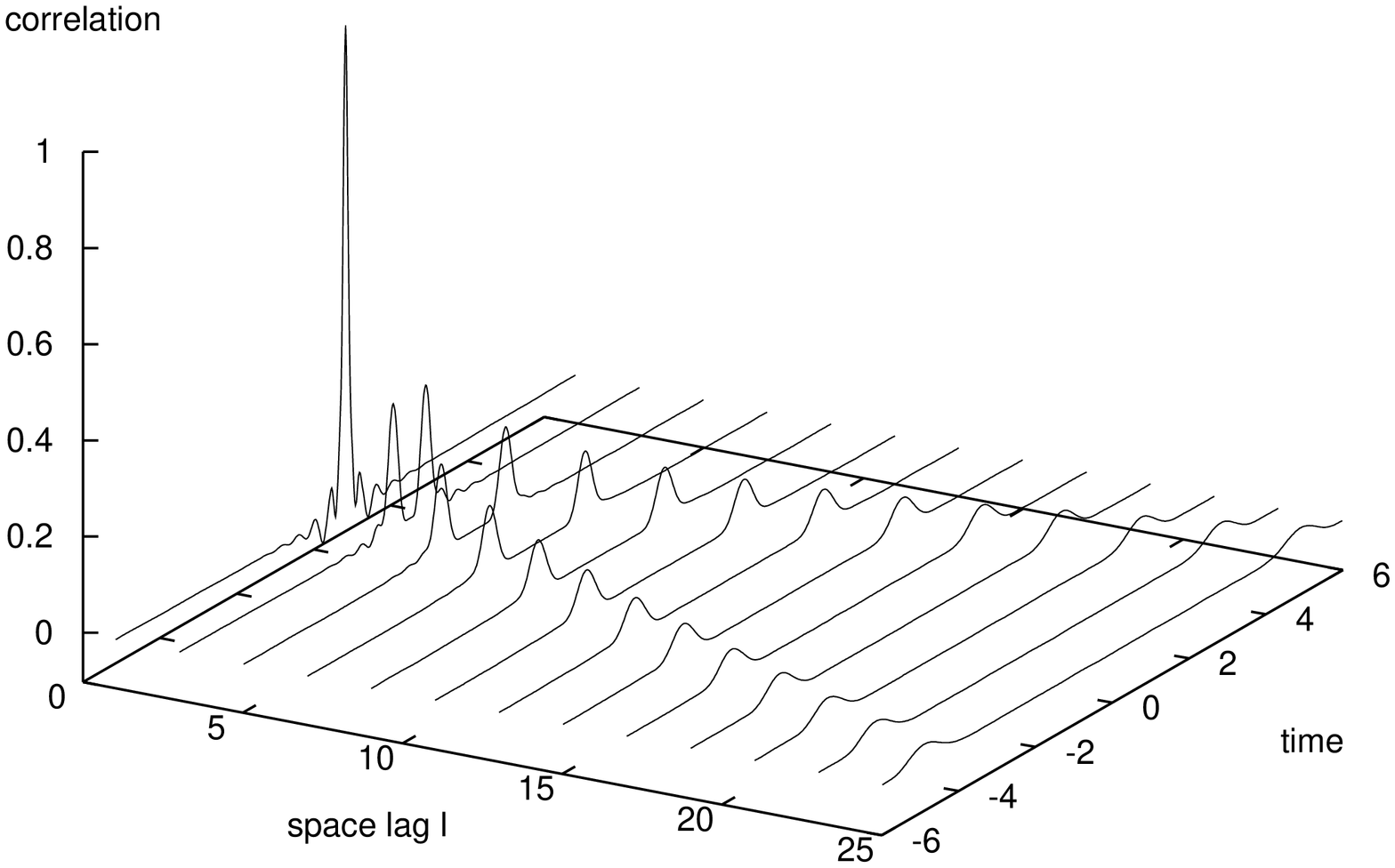,width=12cm,angle=-90}
\begin{minipage}{15cm}
\caption{\baselineskip=12pt
The (normalized) correlation $C_j(l,t)$ of the local flux for the FPU
model with $N=1024$ (same parameters as Fig.~1).
}
\end{minipage}
\end{figure}


\begin{thebibliography}{99}
\bibitem{note1} The statistical independence of the modes stems from
translational invariance, as it can be shown, by computing, e.g., 
$\langle A_k(t)B_{k'}(0)\rangle$ from Eq. (\ref{modi}).
\bibitem{thanks} We are indebted to an anonymous referee for having 
pointed out this connection.
\end{thebibliography}
\end{document}